\documentclass[pra,aps,amsmath,twocolumn,showpacs,floatfix]{revtex4-1}
\usepackage{graphicx}
\usepackage{bm}
\usepackage{pdfpages}
\input{epsf}
\begin{document}
\epsfverbosetrue
\def\la{{\langle}}
\def\ra{{\rangle}}
\def\vep{{\varepsilon}}
\newcommand{\beq}{\begin{equation}}
\newcommand{\eeq}{\end{equation}}
\newcommand{\beqa}{\begin{eqnarray}}
\newcommand{\eeqa}{\end{eqnarray}}
\newcommand{\q}{\quad}
\newcommand{\A}{\alpha}
\newcommand{\h}{\hat{H}}
\newcommand{\ha}{\hat{h}}
\newcommand{\p}{\mbox{p}}
\newcommand{\dg}{+}
\newcommand{\dga}{\dagger}
\newcommand{\AC}{{\it AC}}
\newcommand{\n}{\\ \nonumber}
\newcommand{\om}{\omega}
\newcommand{\Om}{\Omega}
\newcommand{\os}[1]{#1_{\hbox{\scriptsize {osc}}}}
\newcommand{\cn}[1]{#1_{\hbox{\scriptsize{con}}}}
\newcommand{\sy}[1]{#1_{\hbox{\scriptsize{sys}}}}

\title{Interference effects in tunnelling of "cat" wave packet states}
\author {D. Sokolovski$^{a,b}$}
\affiliation{$^a$ Departamento de Qu\'imica-F\'isica, Universidad del Pa\' is Vasco, UPV/EHU, Leioa, Spain}
\affiliation{$^b$ IKERBASQUE, Basque Foundation for Science, Maria Diaz de Haro 3, 48013, Bilbao, Spain}
\begin{abstract}
We analyse tunnelling of a single particle, whose initial state is given by a superposition of spatially 
separated wave packet modes. It is shown that "pile up" of different components the scatterer may change the tunnelling probabilities, making such states a convenient tool for probing the barrier's scattering times. Interference effects arising in  resonance tunnelling in are studied in detail. The analysis allows us to gain further insight into the origin of interference effects in scattering of several identical particles.
\end{abstract}
\pacs{37.10.Gh, 03.75.Kk, 05.30.Jp}
\maketitle

%
%
%
%
%
\section{Introduction}
Recently, there has been considerable interest in the interference effects accompanying scattering of several non-interacting identical particles \cite{INT1}-\cite{DS}. In the celebrated Hong-Ou-Mandel setup (see, for example, \cite{INT3}, \cite{NAT}), such effects are observed if the particles, incident from opposite sides, coincide in the scatterer. Recently it was shown that a different kind of interference effects may arise in the case, where identical particles, incident from the same side, and detained in the scatterer, coincide there due to a kind of "pile up" effect \cite{Baskin}, \cite{SSB}. As a result, in resonance tunnelling, the transmission probability was shown to oscillate as a function of the temporal delay between the arriving particles, provided two or more metastable states in the barrier can be accessed \cite{Baskin}. 
\newline
In this paper, we discuss a closely related case, where the state of a {\it single} particle, incident on a barrier, consists of several spatially separated wave packet modes. Such exotic 'cat' state can be created, for example, by splitting the original wave packet into parts, which experience different time delays before being recombined  \cite{ANN} or, in the case of cold atoms, by using techniques similar to those described in \cite{LCAT}. We will show that the "pile up" of the modes, caused by a delay in the barrier region, can cause observable changes the tunnelling probability. One purpose of this paper is to analyse the use of such systems as an alternative  tool for probing the barrier's scattering times. Its other purpose is to use this analysis in order to gain further
insight into the nature of interference effects in scattering of several identical particles.
\newline
The rest of the paper is organised as follows. In Sect. I we consider transmission of a multi-component initial state. In Sect. II we analyse its transmission across a rectangular barrier. In Sect. III we study the case of resonance tunnelling, and the interference patterns occurring in the transmission probability. In Sect. IV we discuss the interference mechanism, and its similarity with the case of several identical particles. Sect. V consider transmission of a mixed 'cat' state, and Sect. VI contains our conclusions. 
\section{A multi-component initial state}
Consider, in one dimension, a particle whose wave function is given by a superposition of $N$ wave packets, delayed in time relative to each other ($\hbar=1$), 
$\psi_n(x)$,
\begin{eqnarray}\label{1}
\Psi_0(x,t)= K^{-1/2}\sum_{n=1}^N \psi_n(x,t),
\end{eqnarray}
\begin{eqnarray}\label{2}
\nonumber
\psi_n(x,t)=(2\pi)^{-1/2}\int A_n(p) \exp[ipx-iE(p)(t+t_n)]dp, 
\end{eqnarray}
where  $0=t_1<t_2 < ...< t_N$. 
 For a particle of mass $\mu$, e.g., for cold atoms \cite{Baskin} or photons in a waveguide \cite{WG}, the energy is quadratic in the momentum, $E(p)=p^2/2\mu$. For massless particles, e.g., free photons, or electrons in graphene \cite{Khan}, this relation is linear, $E(p)=cp$. The constituent wave packets may, or may not overlap, and for the normalisation constant $K$ in (\ref{1}) we have
 \begin{eqnarray}\label{3}
K=\sum_{mn}\la \psi_m|\psi_n\ra=\q\q\q\q\q\q\q\q\q\q\n
\int dp A_m^*(p)A_n(p)\exp[iE(p)\tau_{mn}]\equiv \sum_{m,n}I_{mn},\q
\end{eqnarray}
where $\tau_{mn}=t_m-t_n$.
The particle is incident on a finite width potential barrier with a transmission amplitude $T(p)$ (see Fig.1) and,
 \begin{figure}
	\centering
		\includegraphics[width=8 cm,height=3cm]{{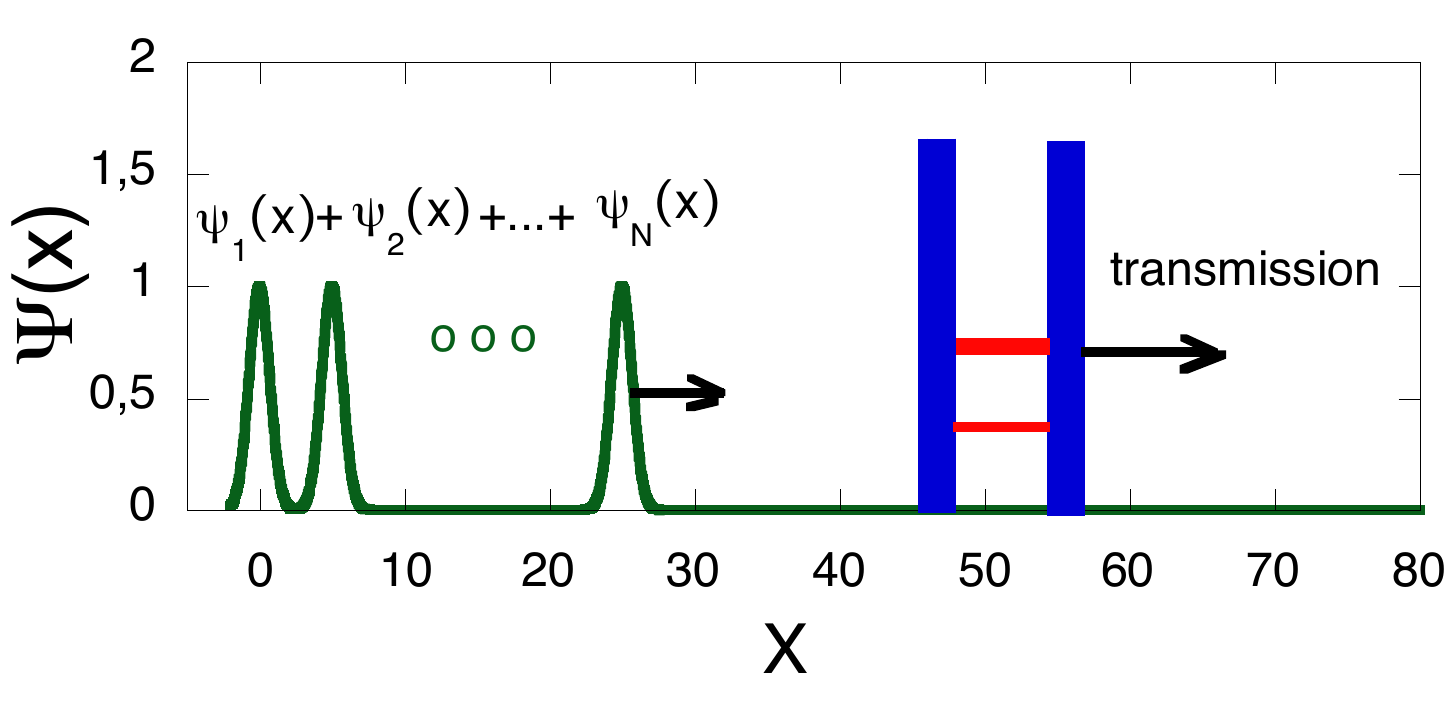}}
\caption{(Color online) Schematic diagram showing a 'cat' state, consisting of $N$ non-overlapping components, incident on a resonance barrier supporting two metastable states}
\label{fig:3}
\end{figure}
 as $t\to \infty$, its  transmitted part takes the form
\begin{eqnarray}\label{4}
\Psi^T(x,t)\equiv \sum_{n}\psi_n^T(x,t)=(2\pi K)^{-1/2}\times \q\q \n
\sum_{n}\int T(p) A_n(p) \exp[ipx-iE(p)(t+t_n)]dp,
\end{eqnarray} 
From Eq.(\ref{4}) the transmission probability we have 
\begin{eqnarray}\label{5}
P^T=\sum_{m,n} T_{mn}/\sum_{m,n} I_{mn},
\end{eqnarray}
where 
\begin{eqnarray}\label{6}
T_{mn}=\la \psi^T_m|\psi^T_n\ra=\q\q\q\q\q\q\q\q\q\q\n
\int dp |T(p)|^2 A_m^*(p)A_n(p)\exp[iE(p)\tau_{mn}],
\end{eqnarray}
is the matrix of the overlaps between the transmitted components $\psi_n^T$.
\newline
If the delays are large, $|t_m-t_n|\to \infty$, rapid oscillations of the exponentials in Eqs.(\ref{3}) and (\ref{6}), make off-diagonal
elements of the overlap matrix vanish, $I_{mn}=a_n\delta_{mn}$ and $T_{mn}=\int|T(p)^2|A_n(p)|^2dp\times \delta_{mn} \equiv w_n\delta_{mn}$. If so, all components of $\Psi_0$ are transmitted independently, and we have 
\begin{eqnarray}\label{7} 
P^T= \sum_n w_n\equiv P^T_{ind}. 
\end{eqnarray}
\newline 
As in \cite{Baskin}, we are interested in the case where $I_{mn}=a_n\delta_{mn}$, and $T_{mn}\ne w_n\delta_{mn}$.
This would indicate that initially non-overlapping components of $\Psi_0$ "pile up" in the barrier region, and the  interference between them affects the outcome of the tunnelling process. A deviation of $P^T$ from $P^T_{ind}$ may, therefore, serve as a crude indicator that a scattered particle spends in the barrier a duration comparable to at least some of $|t_m-t_n|$. 
Next we apply this test to the case of a rectangular barrier.
\section{A rectangular barrier}
For  a rectangular barrier, $V(x)=V$ for $a\le x \le <b$, and $0$ otherwise, 
the transmission coefficient in the tunnelling regime $E(p)<V$ is given by the well known expression 
\begin{eqnarray}\label{y1} 
 |T(p)|^2=1/\{1+V^2 \sinh^2[q(b-a)]/4V(V-E(p)\},\q
\end{eqnarray}
where $q(p)=[2\mu(V-E)]^{1/2}$ for a massive particle. Henceforth, we will consider Gaussian wave packets with
identical momentum distributions, separated by equal time delays,
\begin{eqnarray}\label{y1a} 
A_n(P)=A_m(p)\equiv A(p), \q t_{n+1}-t_n\equiv \tau,\n
|A(p)|^2=(2\pi)^{-1/2}\sigma\exp[-(p-p_0)^2\sigma^2/2].
\end{eqnarray}

 In the deep tunnelling regime, 
$ |T(p)|^2\sim \exp[-2q(b-a)]$ rapidly grows as $E$ increases, but contains no sharp features. As a result, the momentum distribution of each $\psi^T_n$ is shifted towards higher $p$'s, but not modified sufficiently to prevent integrals in Eq.(\ref{6}) from being destroyed by the oscillations (see inset in Fig. 2). There is, therefore, no evidence that different components of the wave function, well separated initially, may be delayed, and eventually "meet"  in the barrier region.
The transmission probability for an $2$- and $5$-component states is shown in Fig.2. As expected, as soon as the overlap between different  $\psi_n$ vanishes, different modes in Eq.(\ref{1})  tunnel independently, and we have $P^T=P^T_{ind}$. 
The absence of the said pile up effect is consistent with the original McColl's suggestion that there is no appreciable delay in tunnelling across a rectangular barrier \cite{MCOLL}. It is also consistent with the finding of Ref.\cite{Baskin}, where tunnelling of two identical particles was studied in a similar context.
\begin{figure}
	\centering
		\includegraphics[width=8 cm,height=4cm]{{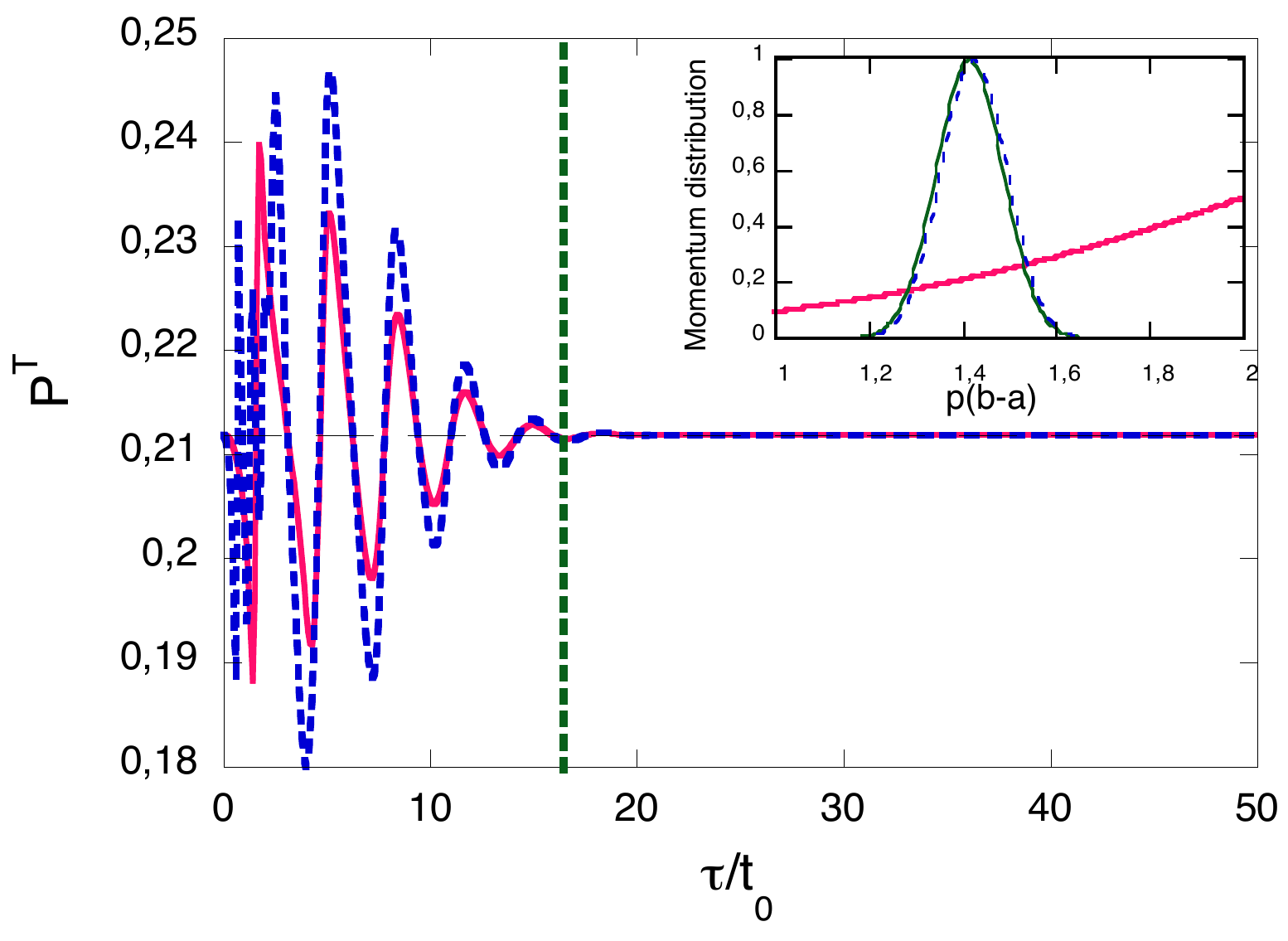}}
\caption{(Color online) Tunnelling probabilities for the rectangular barrier (\ref{y1}) vs. $\tau/t_0$, $t_0\equiv1/2\mu (b-a)^2$, with $2\mu V(b-a)^2=4$. Shown are the cases of $N=2$ (solid) and $N=5$ (dashed) Gaussian components (\ ref{y1a}) with $p_0(b-a)=1.41$, and $\sigma/(b-a)=4.47$. To the right the vertical dashed line the overlap between the components, $\sum_{i\ne j}^N|I_{ij}|$,  is less than $0.005$. Also shown in the inset are $|A(p)|^2$ (solid) and  $|T(p)|^2|A(p)|^2$ (dashed), both renormalised to unit heights, as well as $|T(p)|$ (thick solid).   }

\label{fig:1}
\end{figure}
\section{Resonance tunnelling}
The situation is different in resonance tunnelling across a symmetrical barrier. 
The transmission coefficient of such barrier typically exhibits well separated sharp narrow peaks which, in the Breit-Wigner approximation,  have a Lorentzian shapes,
\begin{eqnarray}\label{a1} 
|T(p)|^2\approx \sum_{j} \frac{\Gamma_j^2}{(p^2/2\mu-E^r_j)^2+\Gamma_j^2}.
\end{eqnarray}
Thus, $E_j^r=E(p^r_j)$ gives the position of the $j$-th resonance peak, and $\Gamma_j$ is its width, and the transmitted momentum distribution, $|T(p)|^2|A(p)|^2|$, can be made much narrower than the incident one, $|A(p)|^2|$.
Approximating both  $|A(p)|^2$ and $\partial_pE$ constant for $E(p)\approx E_j^r$,
and evaluating the remaining integrals in (\ref{6}), then yields
\begin{eqnarray}\label{a2}
T_{mn}=\sum_j C_j\exp(-\Gamma_j|m-n|\tau) \exp[iE^r_j(m-n)\tau]\q\q
\end{eqnarray}
where $C_j= 2\pi \Gamma_j|A(p^r_j)|^2/\partial_pE(p_j^r)$.
Eq. (\ref{a2}) shows that $T_{mn}(\tau)$ oscillates with the internal frequencies of the resonance barrier, $\omega_j=E_j$, $j=0,1,2,...$. In particular, for $N=2$, and just one resonance state at $E^r_1$, we have
the interference correction $\delta P^T(\tau) \equiv P^T(\tau)-P^T_{ind}(\tau)$ is given by ($\tau\equiv \tau_{12}$)
\begin{eqnarray}\label{b2} 
 \delta P^T(\tau)=  \frac{2\mu\pi\Gamma_1}{\partial_pE(p_1^r)}|A(p^r_1)|^2\exp(-\Gamma_1\tau)\cos(E^r_1\tau).\q
\end{eqnarray}
For a narrow resonance,
$ \delta P^T(\tau)$ in Eq.(\ref{b2}) oscillates with a frequency $E_1^r$, persists for the delays at which  $\psi_1$ and $\psi_2$ no longer overlap,  and finally vanishes for $\tau$ exceeding $1/\Gamma_1$,
as illustrated in Fig. 3a. 
\begin{figure}
	\centering
		\includegraphics[width=8 cm,height=4.5cm]{{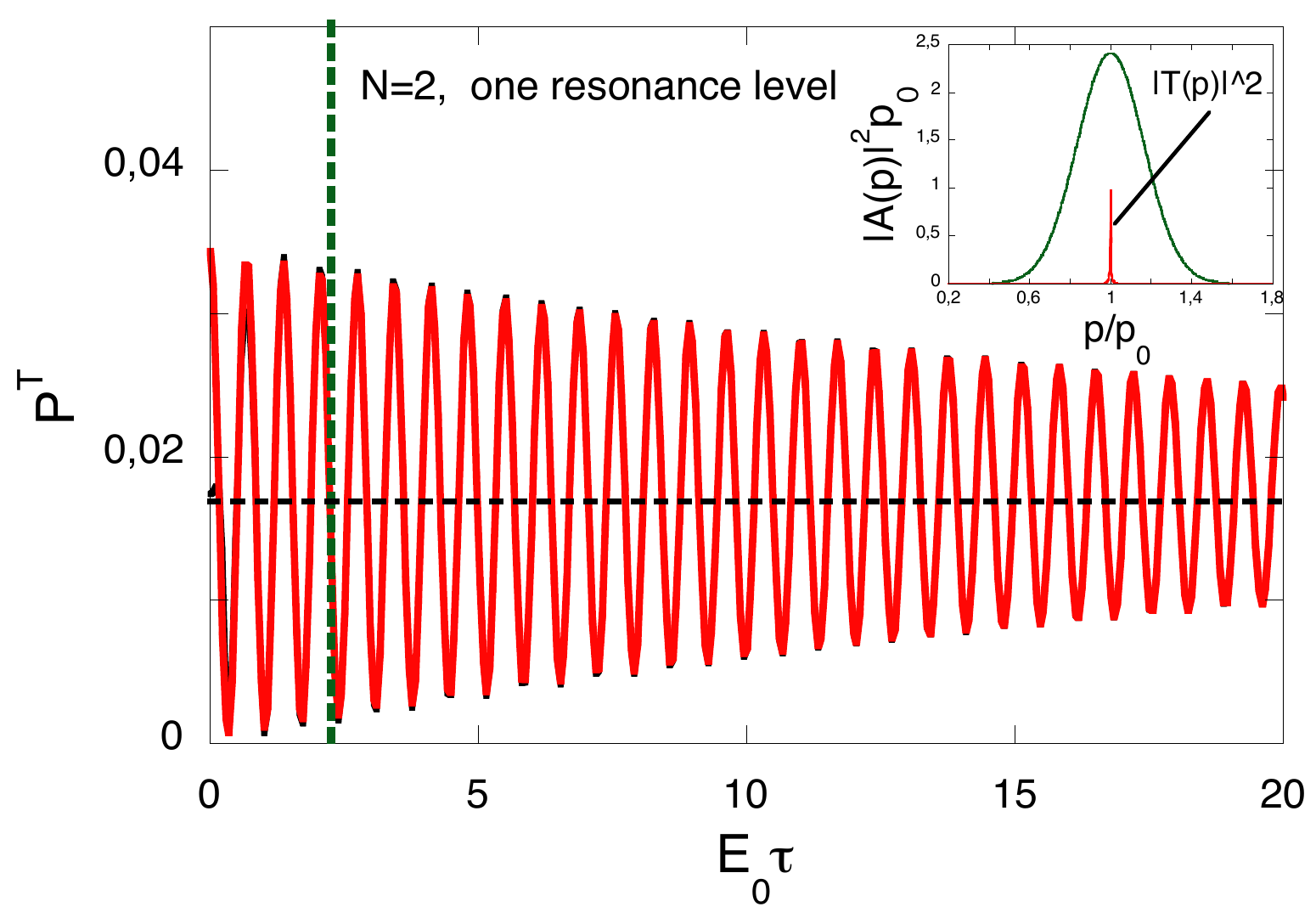}}
\caption{(Color online) Probability to tunnel with one resonance level accessible to a particle in 2-component initial state (\ref{y1a}) vs.
the time lag $\tau=t_2-t_1$, 
for $p_0\sigma=6$, $E_1^r/E_0=1$, $\Gamma_1/E_0=0.014$ and $E_0\equiv E(p_0)$. A horizontal dashed line marks $P^T_{ind}$ in Eq.(\ref{7}); $\psi_1$ and $\psi_2$ can be considered non-overlapping to the right of the vertical dashed line. The inset shows $|A(p)|^2$ and $|T(p)|^2$}
\label{fig:2}
\end{figure}
This is an agreement with the broadly accepted view that, in resonance tunnelling, a 
particle spends approximately a duration of order of the life time of the metastable state
supported by the barrier(see also \cite{Baskin}).
\newline
With only two resonances accessible to the incident particle,
for   $C_1\approx C_2=C$, and $|\tau( \Gamma_{1}-\Gamma_2)|<<1$, we find
\begin{eqnarray}\label{b4} 
 \delta P^T(\tau)\approx
  2C\exp(-\Gamma_1\tau)
  \cos(\delta \omega \tau)\cos(\overline{\omega}\tau),\q\q
\end{eqnarray}
where $\overline{\omega}=(E^r_1+E^r_2)/2$, and $\delta \omega=(E^r_1+E^r_2)/2$.
\begin{figure}
	\centering
		\includegraphics[width=8 cm,height=4.5cm]{{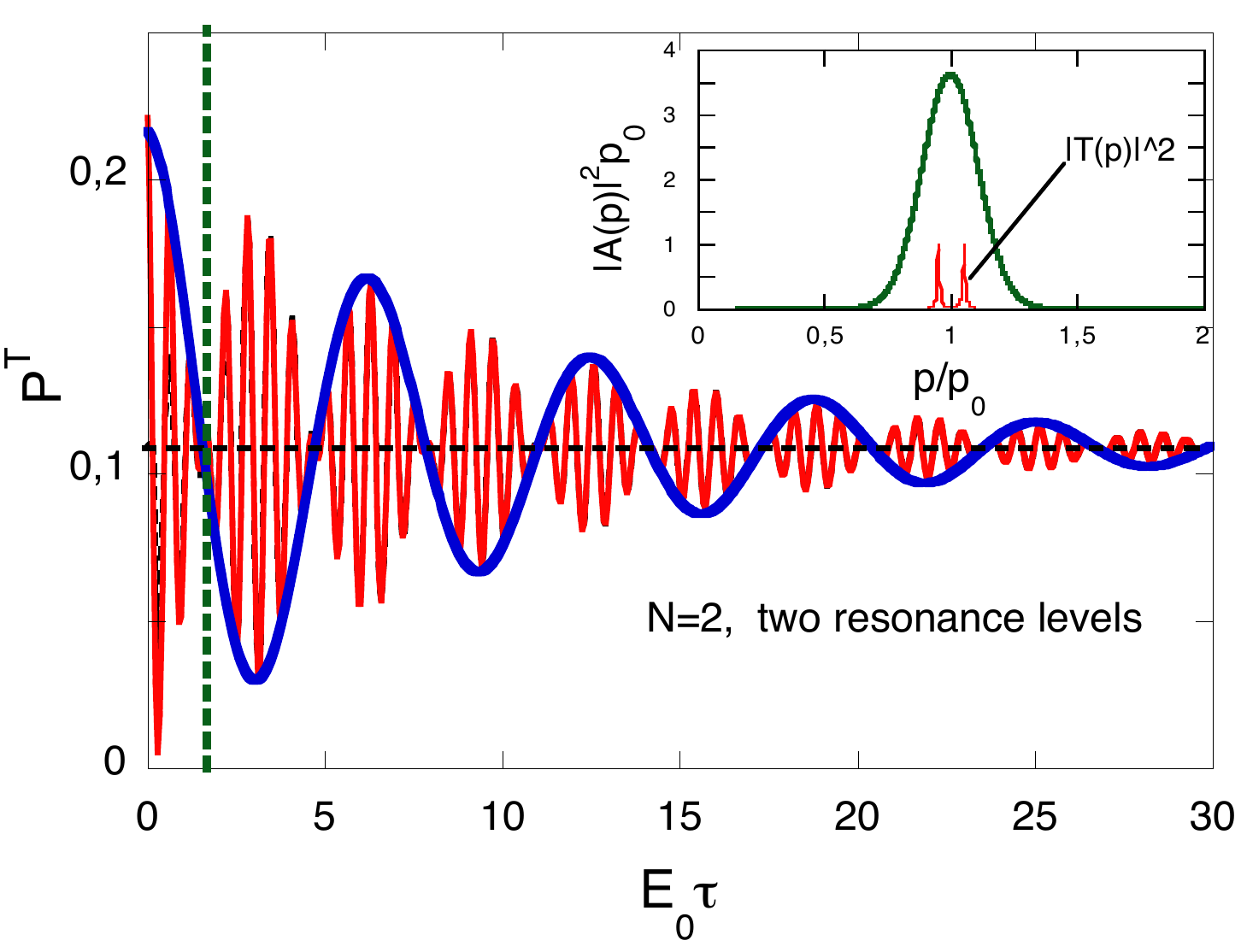}}
\caption{(Color online) Same as Fig.3, but for two resonance levels, $p_0\sigma=9$, $E_1^r/E_0=0.9$, $\Gamma_1/E_0=0.032$, 
$E_1^r/E_0=1,1$, and  $\Gamma_1/E_0=0.038$. Also shown by a thick solid line is the envelope $2C\exp(-\Gamma_1\tau)\cos(\delta \omega \tau)$ in Eq.(\ref{b4}).}
\label{fig:3}
\end{figure}
If two resonance levels are close to each other,  $ \overline{\omega}>>\delta \omega$, damped rapid oscillations of $\delta P(\tau)$ are modulated 
with a much lower frequency $\delta \omega$ (see Fig. 3b). We recall that for two identical particles, quantum statistical correction to transmission probability was found to oscillate with the frequency $2\delta\omega$ \cite{Baskin}, which suggests certain similarity between two effects. We will return to discuss it in Section VI.  
\newline
Finally, for an initial state (\ref{y1a}), containing $N$ identical modes, we have
\begin{eqnarray}\label{e1} 
P^T(\tau)\equiv Nw+\sum_jF_j(\tau,E^r_j,\Gamma_j), 
\end{eqnarray}
where the function $F_N(E^r,\Gamma)$ is given by
\begin{eqnarray}\label{ap2} 
 F_j(\tau,E^r,\Gamma)=C_j Re\{\frac{N}{\exp(-i\mathcal{E}_j^r\tau)-1}-\q\q\q\q\q\q\q\n
\frac{\exp[i\mathcal{E}_j^r(N-1)\tau]-\exp(-i\mathcal{E}_j^r\tau)}{[\exp(-i\mathcal{E}_j^r\tau)-1]^2}\},\q\q
\end{eqnarray}
and we have introduced complex energies $\mathcal{E}_j^r=E_j^r-i\Gamma_j$ to shorten the notations.
For $N>>1$, it is sufficient to retain only the first term in the curly brackets, which yields
\begin{eqnarray}\label{ap3} 
F_N(\tau,E^r,\Gamma)=-CN \frac{\cos(E^r\tau)-\exp(-\Gamma \tau)}{\cos(E^r\tau)-\cosh(\Gamma \tau)}\q\q\q\q\q\q\q\q\q
\end{eqnarray}
Thus, for $\Gamma\tau<<1$, $F_N(\tau,E^r,\Gamma))$ has sharp peaks at $\tau=2\pi k/E^r$, $k=1,2,...$, whose heights are proportional 
to $NE^r/k\Gamma$. When added together, the peaks may give $\delta P^T$ a highly irregular shape, as shown in Fig. 5.
\begin{figure}
	\centering
		\includegraphics[width=8 cm,height=4.5cm]{{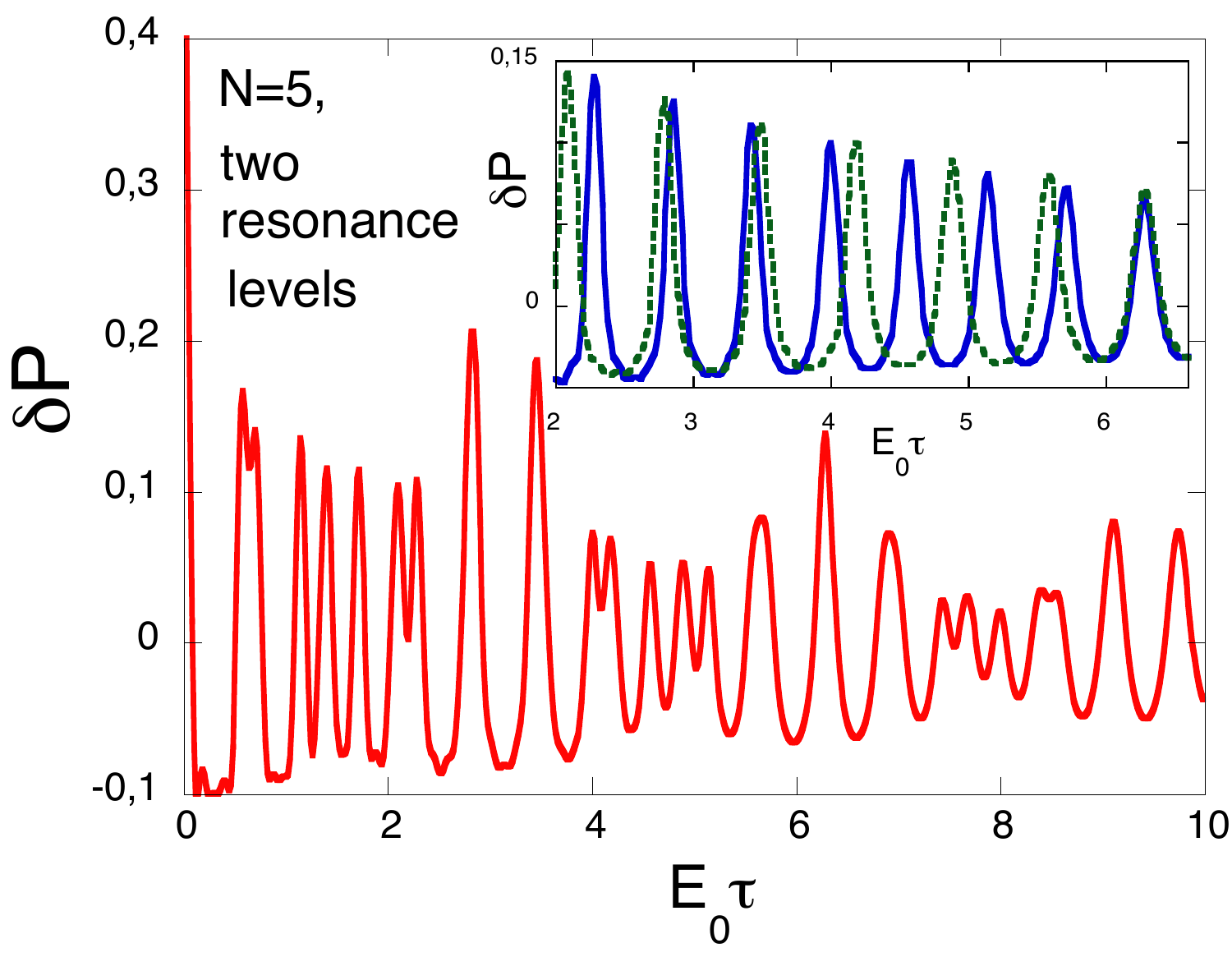}}
\caption{(Color online) Same as Fig. 4, but for five identical Gaussian components, $N=5$, and $t_{n}-t_{n-1}=\tau$. The inset shows 
the sequences of peaks [cf. (\ref{ap2})], contributed by each of the two resonances.}
\label{fig:3}
\end{figure}
\section{The interference mechanism}
The task of evaluating $P^T(\tau)$ is particularly straightforward, since we need not follow the evolution of the wave function in the barrier region, an only require the final overlap matrix $T_{mn}$. By filtering momenta of a $\psi_n$, a narrow resonance at $E^r_j$ produces a nearly monochromatic transmitted state $\psi^T_{nj}$ with $E\approx E_j^r$. The state is broad in the coordinate space, and the overlaps between $\psi^T_{mj}$ and $\psi^T_{nj}$, whose relative delay is $(m-n)\tau$, contains a factor $exp[iE^r_j(m-n)\tau]$. For several well-separated resonances, $\psi^T_{nj}$ and $\psi^T_{nj'}$ with $j\ne j'$ have different energies, and are practically orthogonal,
so that transmissions via different metastable states are mutually exclusive events.
In this way, summation over all summation over all transmitted modes in Eq.(\ref{6}) produces interference structures shown in Figs. 3-5.
\newline
We illustrate this with a simple example, by considering non-spreading wave packet states with a linear dispersion law, $E(p)=cp$.  Assuming the Breit-Wigner form for the transmission amplitude, $T(p)=i\Gamma/[(E-E^r)+i\Gamma]$, and putting $A(p)\approx A(p^r)$, we have 
\begin{eqnarray}\label{2m}
\psi^T_n(x,t)\approx
\frac{2\pi \Gamma A(p^r)}{c}
\Phi(x-ct-ct_n) \q\q
\end{eqnarray}
where
\begin{eqnarray}\label{3m}
\Phi(y)\equiv \theta(-y)\exp(ip^ry+\Gamma y/c),\q\q
\end{eqnarray} 
and $\theta(y)=1$ for $y\ge0$ and $0$ otherwise. Thus, $\psi^T_n$ in Fig.6 has a sharp front 
 followed by a long exponential tail, which allows $\psi_n^T$ to overlap, $\la\psi_m|\psi_n\ra\sim \exp[-iE^r(t_m-t_n)]$, even if $\psi_n$ didn't, $I_{mn}=0$ for $m\ne n$.
  With two, or more, resonances involved $\psi^T_n$ would contain
several contributions of the form (\ref{2m}), one for each metastable state.
\begin{figure}
	\centering
		\includegraphics[width=8 cm,height=4cm]{{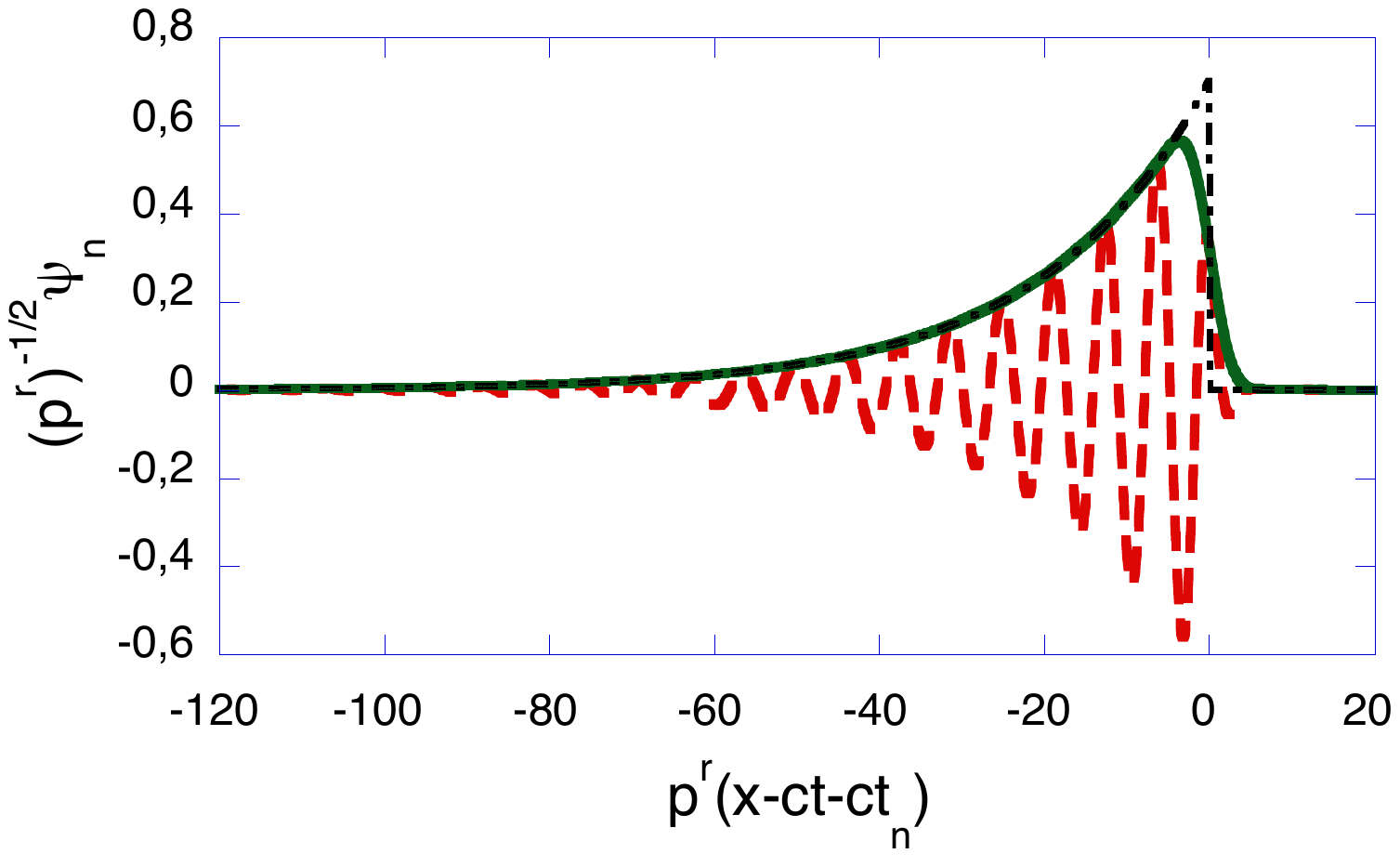}}
\caption{(Color online) Real part (dashed) and modulus (solid) of $\psi_n^T$ in Eq.(\ref{2m}) transmitted across a barrier supporting a single metastable state, with $p^r\Gamma/c=0.05$ and $A(p^r)\Gamma/c=0.11$. Also shown by the dot-dashed line is the modulus of $\psi_n^T$ as give by Eq.(\ref{2m})}.
\label{fig:3}
\end{figure}
\newline
Our analysis of a {\it single} particle, prepared in an exotic initial state, helps us to gain an insight into the interference effects accompanying scattering of {\it several} identical particles. Recently we considered \cite{Baskin} the case of two fermions or bosons, emitted in the same wave packet state, with a time delay $\tau$ between the emissions.
For a barrier with two resonance levels, the 2-particle transmission probability $P^T(2,2)$, considered as a function of $\tau$, exhibited oscillations with a frequency $\Delta \omega=E_2^r-E_1^r$. The oscillations disappear if the particles can be distinguished.
\newline
We note that in both problems we only require the initial and final overlap matrices, $I_{mn}$, and $T_{mn}$ \cite{Baskin}. They are, however, used differently. Whereas in the single particle case, considered here, the correction $\delta P^T(\tau)= 2Re T_{12}$ contains all barrier frequencies, $E_j^r$, the correction to $P^T(2,2)$ depends on $|T_{12}|^2$, and oscillates only with  $\Delta \omega$
\newline
What makes the two cases similar, is that the particle, or particles, are distributed between wave packet modes 
$\psi_1$ and $\psi_2$. For a single particle, this is readily seen from Eq.(\ref{1}).
 A  symmetrised (antisymmetrised) state of two uncorrelated particles, $I_{12}=I_{21}=0$ is given by 
$\Psi(x_1,x_2,t) =
[\psi_1(x_1,t)\psi_2(x_2,t)\pm\psi_1(x_2,t)\psi_2(x_1,t)]/\sqrt{2}$,
which also implies that particle $1$ is simultaneously present in both $\psi_1$ and $\psi_2$, albeit in a different manner.
In both cases, the physical origin of an oscillatory pattern in the transmission probability  is the overlap between different $\psi^T_n$, 
acquired in the barrier, and the phases carried by different $\psi_n$, launched at different times.
\section{Transmission of a mixed cat state}
Before concluding, we briefly discuss transmission of a mixed cat state with two components, 
$ \psi_1(x,t)$ and $ \psi_2(x,t)$, $\la \psi_m|\psi_m\ra=\delta_{mn}$, $m,n=1,2$. The system is prepared in the following way: with a probability $\p/2$ it is in one of  the states 
$ \psi_1(x,t)$ and $ \psi_2(x,t)$, and with a probability $(1-\p)$ it is in their coherent superposition, $ [\psi_1(x,t)+ \psi_2(x,t)]/\sqrt{2}$.
The incident density matrix is, therefore, given by
\begin{eqnarray}\label{f1}
\nonumber 
\rho(x,x')=[\psi_1(x,t)\psi^*_1(x',t)+
\psi_2(x,t)\psi^*_2(x',t)]/2+\\
(1-\p)[\psi_1(x',t)\psi^*_2(x,t)+\psi_2(x,t)\psi^*_1(x',t)]/2. \q\q
\end{eqnarray}
For the transmission probability we have
\begin{eqnarray}\label{f2} 
P^T=[w_1+w_2]/2+(1-\p)Re[T_{12}(\tau)]
\end{eqnarray}
Thus, for a pure state, $\p=0$ we recover Eq.(\ref{5}). As $\p$ increases, the last interference term 
in Eq.(\ref{f2}) becomes smaller, and finally vanishes for the incoherent combination of the two states, $\p=1$, 
where $P^T=(w_1+w_2)/2$. This simple result is easily extended to the case where the initial state has three, or more, components, 
$N>2$.
\section{Conclusions and discussion}
In summary, various components of the same one-particle state, well separated initially, may coincide inside a scatterer, provided the
particle is detained there for an appreciable period of time. Then the interference, resulting from this "pile up" effect, may significantly change the tunnelling probability $P^T$. The effect requires that transmitted modes are significantly broadened in the coordinate space, and is absent in tunnelling across a rectangular barrier. It is present in resonance tunnelling, where $P^T$, considered as a function of temporal delays between the components, oscillates with internal frequencies of the barrier. We have shown that the interference patterns predicted for the resonance transmission of several identical particles \cite{Baskin}, \cite{SSB}, have a similar origin, both resulting from the particle being distributed, in one way or another, between different wave packet components. 
The interference patterns are washed out if the initial state is mixed, rather than pure. The proposed type of "interferometry in the time domain"
is within capability of modern experimental techniques \cite{LCAT},\cite{LAND}.
\newline Support of the Basque Government (Grant No. IT-472-10), of the Ministry of Science and Innovation of Spain (Grant No. FIS2009-12773-C02-01), and useful duscussions with J. Siewert are gratefully acknowledged.

 \end {document}